\documentclass[final,3p,times,twocolumn]{elsarticle}

\usepackage{epsfig}

\biboptions{}

\journal{Physics Letters B}

\begin{document}

\begin{frontmatter}


\author{B.A. Arbuzov}
\ead{arbuzov@theory.sinp.msu.ru}
\author{I.V. Zaitsev}
\address{Lomonosov Moscow State University, Leninskie gory 1, 119991 Moscow, RF}

\title{LHC diboson excesses as an evidence for a heavy WW resonance}




\begin{abstract}
Recently reported diboson excesses at LHC are interpreted  to be connected with heavy $WW$ resonance with weak isotopic spin 2. The resonance appears
due to the wouldbe anomalous triple interaction of the weak bosons, which is defined by well-known coupling constant $\lambda$. We obtain estimates for the effect, which qualitatively agree with ATLAS data. Effects are predicted in
inclusive production of $W^+W^+, W^+ (Z,\gamma), W^+ W^-, (Z,\gamma) (Z,\gamma), W^- (Z,\gamma), W^-W^-$
resonances with $M_R \simeq 2\,TeV$, which
could be reliably checked at the upgraded LHC with $\sqrt{s}\,=\,13 TeV$.
In the framework of an approach to the spontaneous generation of of the triple anomalous interaction its coupling constant is estimated to be  $\lambda = -\,0.017\pm 0.005$ in an agreement with existing restrictions.
\end{abstract}


\end{frontmatter}

\section{Heavy WW resonance}
In experiment~\cite{ATLAS} indications for excesses in the production of boson pairs  $WW,\,WZ,\,ZZ$ were observed
at invariant mass $M_R\simeq 2\,TeV$. Data for these processes are also present in works~\cite{CMSE1,CMSE2}. Despite the fact that the wouldbe effect is not finally established yet, the publication causes numerous
proposals for an interpretation mostly in terms of theories beyond the Standard Model~\cite{INTER1,INTER2,INTER3,INTER4,INTER5,INTER6}, for the further
exhaustive list of references see~\cite{INTER7}.

However it would be advisable to consider for the beginning a possibility
for a more
conventional  interpretation. Indeed, pair of triplets $W^a$ could form
a resonance state, the so-called $W$-ball. Of course the well known gauge interaction of these
bosons with coupling $g(M_W)=0.65$ can not bind them in the resonance state
with mass being of a $TeV$ scale. However, there might exist also an additional effective interaction, {\it e.g.} the anomalous triple boson
interaction~\cite{Hag1,Hag2}, which increases with increasing energy
scale.
In case the interaction becomes sufficiently strong at  a $TeV$ scale,
it might lead to a formation of a resonance under discussion. We shall
consider
this possibility in more details below. But firstly we would discuss if
this option could be reconciled
with present data~\cite{ATLAS,CMSE1,CMSE2}.

Let us assume
an existence of a resonance in a system of two $W$-s. There are few options in the framework of the assumption. Boson W has the unity weak isotopic spin.\footnote{
In what follows we omit {\it weak} in respect to an isotopic spin for the sake of a brevity.} Thus a resonance can have one of possible three values of the isotopic spin: $0,\,1,\,2$. There is also a correlation of values of a spin and an isotopic spin. There are the following options for low spin resonances.
\\
1. Resonance $X_1$ with spin 1 and isotopic spin 1. Then it can not have decay channel to $Z\,Z$ and so it does not correspond to data~\cite{ATLAS}.\\
2. Resonances $X_0,\,X_2$ with spin 0, where a subscript indicate a value of  an isotopic spin. The first option $X_0$ is excluded due to the presence in
data~\cite{ATLAS} of the decay channel $X \to WZ$.

Thus we are left with option $X_2 \equiv X_{ab}$ of a scalar resonance
with the isotopic spin two. Under this premise
we have the following effective interaction
\begin{eqnarray}
& &L_{eff}\, = \,\frac{G_X}{4}\, X_{ab}\, W^c_{\mu\nu}\, W^d_{\mu\nu}\times\nonumber\\
& &\Bigl(\frac{1}{2}\,(\delta^a_c\,\delta^b_d+
\delta^a_d\,\delta^b_c)-\frac{1}{3}\,\delta^{a b}\,\delta_{c d}\Bigr)\,;
\label{GXWW}
\end{eqnarray}
and five charge states of $X_{ab}$
\begin{equation}
X^{++},\;X^+,\;X^0,\;X^-,\;X^{--}\,.\label{Xstates}
\end{equation}
Interaction~(\ref{GXWW}) corresponds to the following form of the effective vertices for
different charge states with momenta and indices of $W$-s $(p,\mu),(q,\nu)$
\begin{eqnarray}
& &X^{++}W^-W^-:\;-\imath\,G_X\,(g_{\mu\nu} (pq)-p_\nu q_\mu);\nonumber\\
& &X^+W^-W^0:\;-\imath\,\frac{G_X}{\sqrt{2}}\,(g_{\mu\nu} (pq)-p_\nu q_\mu);\nonumber\\
& &
X^{0}W^+W^-:\;-\imath\,\frac{G_X}{\sqrt{6}}\,(g_{\mu\nu} (pq)-p_\nu q_\mu);
\label{VERT}\\
& &X^{0}W^0W^0:\;-\imath\,G_X\sqrt{\frac{2}{3}}\,(g_{\mu\nu} (pq)-p_\nu q_\mu);
\nonumber\\
& &X^{-}W^+W^0:\;-\imath\,\frac{G_X}{\sqrt{2}}\,(g_{\mu\nu} (pq)-p_\nu q_\mu);\nonumber\\
& &X^{--}W^+W^+:\;-\imath\,G_X\,(g_{\mu\nu} (pq)-p_\nu q_\mu).\nonumber
\end{eqnarray}
For the neutral boson we have as usually the Weinberg mixture of the $Z$-boson and the photon
\begin{equation}
 W^0\,=\,Z\,\cos\theta_W\,+\,A\,\sin\theta_W\,;\label{Weinberg}
\end{equation}
Effective coupling constant $G_X$ is related to width $\Gamma_X$ of the
resonance
\begin{equation}
\frac{G_X^2 \sqrt{M_X^2-4 M_W^2}(M_X^2-4 M_W^2+6 M_W^4/M_X^2)}{64\,\pi}.\label{width}
\end{equation}
Data~\cite{ATLAS} allow to estimate $\Gamma_X$ to be few hundreds
GeV by an order of magnitude . Thus we shall present estimates  of
cross sections for  three values of the width
\begin{eqnarray}
& &\Gamma_X\,=\,200\,GeV;\quad G_X=\frac{0.002253}{GeV}\,;\nonumber\\ & & \Gamma_X\,=\,300\,GeV;\quad G_X=\frac{0.002759}{GeV}\,;\label{widthn}\\ & &\Gamma_X\,=\,400\,GeV;\quad G_X=\frac{0.003186}{GeV}\,.\nonumber
\end{eqnarray}

With values for effective couplings~(\ref{widthn}) and the mass of
the resonance $M_X = 2000\,GeV$ we calculate cross
sections of the resonances production at two energies:
$\sqrt{s}=8\,TeV$ for the
sake of comparison with data~\cite{ATLAS} and $\sqrt{s}=13\,TeV$ in
view of a prediction of possible effects at the upgraded LHC.
Calculations are performed with application of CompHEP
package~\cite{CompHEP} and a necessary information from PDG~\cite{PDG}.
The results are presented in
Tables.

In the last six lines of the Tables we show results with no difference
in sign of $W^{\pm}$ charge and specification of the neutral component
to be either $Z$ or the photon, that could be
confronted with data~\cite{ATLAS}. We see, that cross sections being
around few $fb$ for channels $WW$, $WZ$, $ZZ$ at $\sqrt{s}=8\,TeV$ qualitatively correspond to data collected with integral luminosity $20\,fb^{-1}$. For more definite conclusions one has to wait for hopefully
forthcoming new data. In any case the most promising process here is a
production of same positive sign $W$ pairs.

\begin{center}
\bigskip
Table 1.\\
Cross sections of X states production for $\sqrt{s}=8\,TeV$ at the
LHC and different channels for three values of the total width.
\end{center}

\begin{center}
\begin{tabular}[l]{|||c|||c||c||c|||
} \hline
$\sqrt{s}$&\multicolumn{3}{|c|||}{$8 \, TeV$}\\ \hline
 $\Gamma_X\,GeV$& 200 & 300 & 400 \\
\hline $\sigma(X^{++})\, fb$& 2.25 &  3.36 & 4.49   \\
 \hline $\sigma(X^{+})\, fb$& 1.98 & 2.89 & 4.04
\\
  \hline $\sigma(X^0)\, fb$ & 1.33 & 2.05 & 2.65 \\
\hline $\sigma(X^{-})\, fb$& 0.61 &  0.95 & 1.24   \\
 \hline $\sigma(X^{--})\, fb$& 0.22 & 0.33 & 0.45
 \\
  \hline $\sigma(WW)\, fb$ & 2.91 & 4.37 & 5.82 \\
  \hline $\sigma(WZ)\, fb$ & 2.00 & 2.95 & 4.05 \\
  \hline $\sigma(W \gamma)\, fb$ & 0.60 & 0.89  & 1.22 \\
  \hline $\sigma(ZZ)\, fb$ & 0.53 & 0.81 & 1.05  \\
  \hline $\sigma(Z \gamma)\, fb$ & 0.31 & 0.49 & 0.63 \\
  \hline $\sigma(\gamma \gamma)\, fb$ & 0.047 & 0.073 & 0.094  \\
 \hline \hline
 \end{tabular}
\end{center}

\begin{center}
\bigskip
Table 2.\\
Cross sections of X states production for $\sqrt{s}=13\,TeV$  at the
LHC and different channels for three values of the total width.
\end{center}

\begin{center}
\begin{tabular}[l]{|||c|||c||c||c|||
} \hline
$\sqrt{s}$&\multicolumn{3}{|c|||}{$13 \, TeV$}\\ \hline
 $\Gamma_X\,GeV$& 200 & 300 & 400 \\
\hline $\sigma(X^{++})\, fb$  &17.9 & 26.9 & 35.8 \\
 \hline $\sigma(X^{+})\, fb$
&16.2 & 22.9 & 33.8\\
  \hline $\sigma(X^0)\, fb$  & 11.4 & 18.1 & 23.6 \\
\hline $\sigma(X^{-})\, fb$ & 6.11 & 9.22 & 12.04 \\
 \hline $\sigma(X^{--})\, fb$
&2.60& 3.90 & 5.22 \\
  \hline $\sigma(WW)\, fb$  & 24.3 & 36.8 & 48.9 \\
  \hline $\sigma(WZ)\, fb$  & 17.2 & 24.7 & 35.2\\
  \hline $\sigma(W \gamma)\, fb$  & 5.20 & 7.39 & 10.7\\
  \hline $\sigma(ZZ)\, fb$  & 4.49 & 7.13 & 9.30 \\
  \hline $\sigma(Z \gamma)\, fb$  & 2.70 & 4.29 & 5.59\\
  \hline $\sigma(\gamma \gamma)\, fb$  & 0.41 & 0.64 & 0.84 \\
 \hline \hline
 \end{tabular}
\end{center}

\section{A model for the $WW$ resonance}

Now let us consider a possibility of a heavy resonance
in case of an existence of the anomalous three-boson interaction, which in
conventional notations looks like
\begin{eqnarray}
& &\frac{g\,\lambda}{3!\,M_W^2}\,F\,\epsilon_{abc}\,W_{\mu\nu}^a\,
W_{\nu\rho}^b\,
W_{\rho\mu}^c\,;
\label{FFF}\\
& &W_{\mu\nu}^a\,=\,
\partial_\mu W_\nu^a - \partial_\nu W_\mu^a\,+g\,\epsilon_{abc}
W_\mu^b W_\nu^c\,;\nonumber
\end{eqnarray}
where $g \simeq 0.65$ is the electro-weak coupling.
The best limitations for parameter $\lambda$ read~\cite{PDG}
\begin{equation}
\lambda_\gamma = -\,0.022\pm0.019\,;\quad
 \lambda_Z = -\,0.09\pm0.06\,; \label{lambda1}
\end{equation}
where a subscript denote a neutral boson being involved in the experimental definition of $\lambda$. Sources of results for $\lambda$ see~\cite{LEP,Atlas1,Atlas2,CMS1,CMS2}.
Let us emphasize that
$F \equiv F(p_i)$ in definition~(\ref{FFF}) denotes a form-factor, which is either postulated as
in original works~\cite{Hag1,Hag2} or it is just uniquely defined as
in works on
a spontaneous generation of effective interaction~(\ref{FFF})
~\cite{BAA09,AVZ2}. In any case the form-factor guarantees the effective interaction to act in a limited region of the momentum space. That is
it vanishes for momenta exceeding scale $\Lambda$. We shall see that for the problem under discussion this scale has to be of order of magnitude
of few $TeV$.

Let us present the expression for the form-factor, which was obtained in
the approach~\cite{BAA09,AVZ2} for special kinematics
$p_1\,=\,0,\,p_2\,=\,-\,p_3\,=\,p$
\begin{eqnarray}
& &u < u_0:\nonumber\\
& & F(0,p,-p) = F(u) = \frac{1}{2}\,G^{31}_{15}
\Bigl(\,u|^0_{0,\frac{1}{2},1,-\frac{1}{2},-1}\Bigr)-\nonumber\\
& &\frac{85\, g(\Lambda) \sqrt{2}}{128 \pi}\,
G^{31}_{15}\Bigl(\,u|^\frac{1}{2}_{\frac{1}{2},\frac{1}{2},1,
-\frac{1}{2},-1}\Bigr)+\label{Formfactor}\\
& &C_1\,G^{10}_{04}\Bigl(\,u|_{\frac{1}{2},1,
-\frac{1}{2},-1}\Bigr)+C_2
G^{10}_{15}\Bigl(\,z|_{1,\frac{1}{2},
-\frac{1}{2},-1}\Bigr);\; \nonumber\\
& &u > u_0:\quad F(u) = 0;\,\quad u = \frac{G^2\,x^2}{512\,\pi^2};\nonumber\\
& & x = p^2;\quad u_0=9.6175 = \frac{G^2\Lambda^4}{512\,\pi^2}\,;
\nonumber\\
& & C_1=-0.035096,\,C_2=-0.051104;\nonumber\\
& &g(\Lambda)=0.60366.\nonumber
\end{eqnarray}

Here
$$
G^{mn}_{qp}\Bigl(\,u|^{a_1,..,a_q}_{b_1,..b_p}\Bigr)
$$
is a Meijer function~\cite{be} (see also convenient set of necessary
formulas in book~\cite{ABOOK}).
Interaction constant $G$ in relations~(\ref{Formfactor}) is connected
with conventional definitions in the following way
\begin{equation}
G\,=\,-\,\frac{g\,\lambda}{M_W^2}\,.\label{Glam}
\end{equation}
Calculations were done in the framework of an approximate scheme, which accuracy was estimated to be $\simeq (10 - 15)\%$~\cite{BAA04}. Would-be existence of effective interaction~(\ref{FFF}) leads to important non-perturbative effects in the electro-weak interaction.
In particular, one might expect resonances to appear in the system of
two $W^a$-bosons. A possibility of an appearance of such states (W-balls)
was already discussed, {\it e.g.} in works~\cite{AVZ2,AZ12}.

Let us  consider a Bethe-Salpeter equation for a scalar bound state or a resonance consisting of two $W$. We have already noted, that such state
can have two values of the weak isospin: $I = 0,\,I = 2$.
\begin{figure}
\includegraphics[scale=0.6]{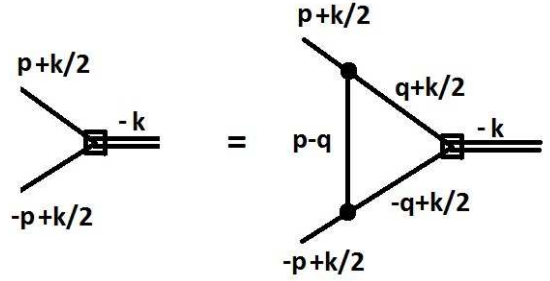}
\caption{Diagram form of equation~(\ref{eq:BS2}). Simple lines represent
$W$-s, a double line corresponds to the resonance, black circles correspond to
interaction~(\ref{FFF}).}
\label{fig:compenG}
\end{figure}

With interaction~(\ref{FFF}) we have the following
equation for a spin zero state in correspondence to diagrams
presented in Fig 1.
\begin{eqnarray}
& & \Psi_R(x) =\frac{G^2 A_I}{8\,\pi^2}\biggl(\int_0^{\Lambda^2} D(y,v) \Psi_R(y) F^2(y)\,y dy +\nonumber\\
& &
\frac{1}{6 x^2}\int_0^x \Psi_R(y) y^3 dy -\frac{1}{3 x}\int_0^x \Psi_R(y) y^2 dy-\nonumber\\
& &\frac{x}{3}\int_x^{\Lambda^2} \Psi_R(y) dy + \frac{x^2}{6}\int_x^{\Lambda^2} \frac{\Psi_R(y)} {y} dy\biggr);\label{eq:BS2}\\
& &D(y,v)=\frac{384 y^3-64 y^2 v+20 y v^2-v^3}{64y^2(4 y+v)};\nonumber
\end{eqnarray}
where $A_I,\,I=0,2$ is an isotopic factor, $x=p^2$ and $y=q^2$, $p$ and $q$ being an external and an internal momentum in Euclidean four-momentum space,
$v=k^2$ with $k^2$ being Euclidean four-momentum squared of the resonance
state. Form-factor F(y) of effective interaction~(\ref{FFF}) is introduced in the constant term of the equation.
Here in view of large value $M_R \simeq 2\,TeV$ of the wouldbe resonance we neglect $W$ mass, and isotopic factor $A_I$ is defined for different values of the total isospin $I$ as follows
\begin{equation}
A_0 = -2;\quad A_2 = 1 .\label{eq:AI}
\end{equation}

In equation~(\ref{eq:BS2}) we have introduced a dependence on momentum $k$
of the bound state only in the first constant term in the RHS part.
This corresponds to approximation which is used in the
approach~\cite{ABOOK}.
The effective cut-off parameter $Y$ in~(\ref{eq:BS2}) is the
same parameter, which was used in works~\cite{BAA09,AVZ2} in the course of
studies of a spontaneous generation of interaction~(\ref{FFF}) and
is defined by relations in~(\ref{Formfactor})

With the following substitution for variables with account
of~(\ref{Formfactor})
\begin{eqnarray}
& &z = \frac{G^2\,x^2}{64\,\pi^2};\; t = \frac{G^2\,y^2}{64\,\pi^2};\; z_0 = \frac{G^2\,\Lambda^4}{64\,\pi^2} =
8\, u_0;\nonumber\\
& &\sqrt{\mu} =
\frac{G\,M_R^2}{8\,\pi}=-\frac{G\,v}{8\,\pi}\label{eq:zy}
\end{eqnarray}
we come to the following equation for the weak isotopic spin $I=2$ \begin{eqnarray}
& & \Psi_R(z) = 4\int_0^{z_0}D(t,\mu)\Psi_R(t) F^2(t) dt +\nonumber\\
& &
\frac{2}{3 z}\int_0^z \Psi_R(t) t dt - \frac{4}{3 \sqrt{z}}\int_0^z \Psi_R(t) \sqrt{t} dt-\nonumber\\
& &\frac{4 \sqrt{z}}{3}\int_z^{z_0} \frac{\Psi_R(t)}{\sqrt{t}} dt +\frac{2 z}{3}
\int_z^{z_0} \frac{\Psi_R(t)} {t} dt;\label{eq:BSZ}\\
& &D(t,\mu)=\frac{384 t\sqrt{t}+64 t \sqrt{\mu}+20\sqrt{t}  \mu+\mu^{3/2}}{64t(4 \sqrt{t}-\sqrt{\mu})}.\nonumber
\end{eqnarray}
Equation~(\ref{eq:BSZ}) being homogenous we choose solution under condition
$\Psi_R(0)=1$, that means
\begin{equation}
B\,=\,4\int_0^{z_0}D(t,\mu)\Psi_R(t) F^2(t) dt\,=\,1.\label{eq:norm}
\end{equation}
By successive differentiations of equation~(\ref{eq:BSZ}) we obtain a Meijer
differential equation for function $\Psi_R(z)$
\begin{eqnarray}
& &\biggl(z\frac{d}{dz}+1\biggr)\biggl(z\frac{d}{dz}+\frac{1}{2}\biggr)
\biggl(z\frac{d}{dz}\biggr)\biggl(z\frac{d}{dz}-\frac{1}{2}\biggr)\times
\nonumber\\
& &
\biggl(z\frac{d}{dz}-1\biggr)\Psi_R(z) + z\,\Psi_R(z)\,=\,0.\label{eq:diff}
\end{eqnarray}
Then we look for the solution in terms of Meijer functions
(see {\it e.g.}~\cite{ABOOK}) in the following form
\begin{eqnarray}
& &\Psi_R(z)\,=\,\frac{1}{2}\,G_{15}^{31}\bigl( \,z\,|\,^0_{0,1/2,1,-1/2,-1}\bigr)+\nonumber\\
& &C_3\,G_{04}^{10}\bigl( z\,|\,^{}_{1,1/2,-1/2,-1}\bigr)+\nonumber\\
& &C_4\,
G_{04}^{10}\bigl( z|\,^{}_{1/2,1,-1/2,-1}\bigr).\label{eq:solC}
\end{eqnarray}
The substitution of~(\ref{eq:solC}) into equation~(\ref{eq:BSZ}) with
condition~(\ref{eq:norm}) gives unique
definition of constants:
\begin{equation}
C_3\,=\,-\,0.0001925,\; C_4\,=\,0.0015612\,.\label{C12}
\end{equation}
Then we substitute solution~(\ref{eq:solC},\ref{C12}) and
form-factor~(\ref{Formfactor}) into normalization
condition~(\ref{eq:norm}) and thus define parameter $\mu$, which according to~(\ref{eq:zy}) is connected with
the mass of the resonance $M_R$.
Let us note, that in relation~(\ref{eq:norm}) one has to calculate the Cauchy principal value of the integral. Namely, for $\mu=0$ in~(\ref{eq:norm})
we have
$B = 0.742411$. With $\mu$ increasing $B$ also increases and $B = 1$ for
\begin{equation}
\mu\,=\,0.072925.\label{eq:mu}
\end{equation}
Then from definitions~(\ref{Glam}) and~(\ref{eq:zy}) we have for $M_R =2\,TeV$
\begin{equation}
\lambda\,=\,-\,0.01687.\label{eq:lam}
\end{equation}
where the sign is defined in the framework of the approach~\cite{BAA09,AVZ2} due to the condition of coupling constant $G$ to be positive. This value agrees limitations~(\ref{lambda1}).
According to relations~(\ref{Formfactor},\ref{Glam}), value~(\ref{eq:lam}) corresponds to the following effective cut-off parameter
\begin{equation}
\Lambda\,=\,(11.4 \pm 1.7)\,TeV\,;\label{LAMBDA}
\end{equation}
where we take into account already mentioned $15\%$ uncertainty inherent to the model. Result~(\ref{LAMBDA}) demonstrates a natural appearance of $TeV$
scale in the model.
With account of the uncertainty in~(\ref{LAMBDA}) our estimate~(\ref{eq:lam}) is to be changed for
\begin{equation}
\lambda\,=\,-\,0.017 \pm 0.005.\label{eq:lam1}
\end{equation}
This result being more reliable than~(\ref{eq:lam}) also is quite compatible with limitations~(\ref{lambda1}).

It is worth noting that for a wouldbe state with isotopic spin zero
according to~(\ref{eq:AI}) there is an overall factor $-2$ in
equation~(\ref{eq:BS2}).
The analogous procedure leads to the conclusion that there is no appropriate
solution of the problem. We conclude, that only isotopic spin two state
may be present in the framework of the present model.

\section{Conclusion}
Thus we consider possible five resonant states with approximate mass $2\,TeV$, which decay into
\begin{eqnarray}
& &W^+W^+;\quad W^+W^0;\quad W^+W^-,W^0W^0;\nonumber\\
& &W^-W^0;\quad W^-W^-.\label{eq:WWZ}
\end{eqnarray}
Here we have to bear in mind relation~(\ref{Weinberg}), which connect field $W^0$ with physical fields of the $Z$ boson
and of the photon. Results of calculations for cross sections of these states
production are shown in Tables 1,2. The results can be reliably checked
in forthcoming experiments at LHC with $\sqrt{s} = 13\,TeV$.

Assuming a spontaneous generation of the anomalous triple weak bosons
interaction~(\ref{FFF}) we predict value $\lambda$~(\ref{eq:lam1}), which
also has an opportunity to be checked at the upgraded LHC.

\section{Acknowledgments}

The work is supported in part by the Russian Ministry of Education and Science
under grant NSh-3042.2014.2.

\end{document}